\documentclass[twocolumn,aps,unsortedaddress,superscriptaddress,prl,floatfix]{revtex4}
\usepackage{graphicx}
\usepackage{subfigure}
\usepackage{pdfpages}
\usepackage{amsmath}

\begin{document}

\title{Materials with low DC magnetic susceptibility for sensitive magnetic measurements}
\author{R. Khatiwada}
\email[Electronic address: ]{rakshyakha1@gmail.com}
\affiliation{Department of Physics, Indiana University, Bloomington, IN, 47405 and IU Center for Exploration of Energy and Matter, Bloomington, IN, 47408, USA}
\affiliation{Department of Physics, Indiana University Purdue University, Indianapolis, IN, 46202, USA}
\affiliation{Department of Physics, University of Washington, Seattle, WA, 98195, USA}
\author{L. Dennis}
\affiliation{Department of Physics, Cornell College, Mt Vernon, IA, 52314, USA}
\affiliation{Department of Physics, University of California Davis, Davis, CA, 95616, USA}
\author{R. Kendrick}
\affiliation{Department of Physics, Indiana University, Bloomington, IN, 47405 and IU Center for Exploration of Energy and Matter, Bloomington, IN, 47408, USA}
\affiliation{Department of Physics, Indiana University Purdue University, Indianapolis, IN, 46202, USA}
\affiliation{Naval Surface Warfare Center Crane Division, Crane, IN, 47522, USA}
\author{M. Khosravi}
\affiliation{Department of Physics, Indiana University, Bloomington, IN, 47405 and IU Center for Exploration of Energy and Matter, Bloomington, IN, 47408, USA}
\affiliation{ProNova Solutions, Maryville, TN, 37804, USA}
\affiliation{Sun Nuclear Corporation, Melbourne, FL, 32940, USA}
\author{M. Peters}
\affiliation{Department of Physics, Indiana University, Bloomington, IN, 47405 and IU Center for Exploration of Energy and Matter, Bloomington, IN, 47408, USA}
\author{E. Smith}
\author{W. M. Snow}
\affiliation{Department of Physics, Indiana University, Bloomington, IN, 47405 and IU Center for Exploration of Energy and Matter, Bloomington, IN, 47408, USA}
\date{\today}

\begin{abstract}
Materials with very low DC magnetic susceptibility have many scientific applications. To our knowledge however, relatively little research has been conducted with the goal to produce a “totally nonmagnetic” material. This phrase in our case means after spatially averaging over macroscopic volumes, it possesses an average zero DC magnetic susceptibility. We report measurements of the DC magnetic susceptibility of three different types of nonmagnetic materials at room temperature: (I) solutions of paramagnetic salts and diamagnetic liquids, (II) liquid gallium-indium alloys and (III) pressed powder mixtures of tungsten and bismuth. The lowest measured magnetic susceptibility among these candidate materials is in the order of $10^{-9}$ cgs volume susceptibility units, about two orders of magnitude smaller than distilled water. In all cases, the measured concentration dependence of the magnetic susceptibility is consistent with that expected for the weighted sum of the susceptibilities of the separate components within experimental error. These results verify the well-known Wiedemann additivity law for the magnetic susceptibility of inert mixtures of materials and thereby realize the ability to produce materials with small but tunable magnetic susceptibility. For our particular scientific application, we are also looking for materials with the largest possible number of neutrons and protons per unit volume. The gallium-indium alloys fabricated and measured in this work possess to our knowledge the smallest ratio of volume magnetic susceptibility to nucleon number density per unit volume for a room temperature liquid, and the tungsten-bismuth pressed powder mixtures possess to our knowledge the smallest ratio of volume magnetic susceptibility to nucleon number density per unit volume for a room temperature solid. This ratio is a figure of merit for a certain class of precision experiments that search for possible exotic spin-dependent forces of Nature.
\end{abstract}

\maketitle

\section{Introduction}
It is often necessary to exploit or invent new procedures and materials which allow one to remove some extraneous component with unwanted physical properties to make progress in experimental physics. Normally this can be done to the required levels of precision only for certain materials with select properties, which thereby become the standard material to use for some particular technique. One example of a precision measurements field which can benefit from such developments is the quest to measure smaller and smaller magnetic fields with higher and higher precision. The precision and the sensitivity of magnetic field measurement have greatly improved over the last few years \cite{Dang:2010, Grith:2010, Patton:2014}. The magnetic susceptibility of the material used in the apparatus places a fundamental limitation on the measurement precision as any nonzero magnetic susceptibility distorts the field one is trying to measure at some level. One therefore wants to develop materials where the susceptibility is matched to some value that affords the least distortion for the measurement of interest. Extensive work has been done in the NMR/MRI community to develop probes ~\cite{Ravi:2010, Tse:2011} and materials~\cite{Doty:1998, Walper:2014} whose magnetic susceptibility match to that of certain tissues so as not to distort the MRI images. Low magnetic susceptibility materials are also of interest in other areas of magnetic instrumentation~\cite{Sunderland:2009} and for applications in standards laboratories in devices such as the watt balance~\cite{Silvestri:2003}. The recently-launched Lisa Pathfinder mission,whose scientific goal is to perform the R\&D needed to develop a space-based measurement system for gravitational waves, employs test masses made of a low magnetic susceptibility gold-platinum alloy in part to minimize magnetic effects ~\cite{Jennrich:2009, Silvestri:2003, Budworth:1990}. There is extensive data both on the magnetic susceptibility of the elements and simple compounds~\cite{Landolt:1986, Landolt:1992} and also on practical materials for use in apparatus construction sensitive to magnetic fields both at room temperature~\cite{Keyser:1989} and at lower temperatures ~\cite{Lockhart:1990}.\\

Our application of interest is to search for possible exotic spin-dependent interactions of nature. Many experimental groups are searching for possible exotic interactions which depend on spin ~\cite{Chu:2013, Tullney:2013, Leslie:2014}. These experiments typically employ one test mass which includes either spin-polarized electrons or nuclei and a second nonmagnetic test mass which is moved close to and far from the polarized ensemble. As the energy shift of the spins from this possible new spin-dependent interaction comes from the sum over the individual protons, neutrons, or electrons in the atoms, it is clear that denser materials for the test mass are preferable. However, all of these spin ensembles are either immersed in an external magnetic field or themselves generate a magnetic field through their magnetic moments. The introduction of a test mass with finite magnetic susceptibility near and far from the ensemble necessarily disturbs the magnetic field and therefore the spin dynamics of the ensemble, thereby generating a systematic error in the experiment. The magnetic field generated by polarized electron sources of nontrivial density are naturally much larger than those from polarized nuclei due to the much larger electron magnetic moment and some work has been done to prepare special materials which can suppress these external fields ~\cite{Ritter:1990, Heckel:2008, Leslie:2014, Ni:1994, Ni:1993, Chui:1993}.\\

The systematic error from magnetic susceptibility threatens to constitute a fundamental limitation on measurements of this type. This issue is not an abstract one: already systematic effects from the magnetic susceptibility of the test masses in these experiments are starting to become a serious problem. In the work of ~\cite{Hoedl:2011} which used a torsion balance to set a limit on possible monopole-dipole interactions involving polarized electrons, it was necessary in the end to coat the silicon test mass with paramagnetic terbium and exploit the variation of its magnetic susceptibility with temperature to cancel a systematic error coming from the finite magnetic susceptibility of silicon. In the work of ~\cite{Tullney:2013}, which used a Bismuth Germanium Oxide (BGO) crystal as a test mass of high nucleon density to search for neutron monopole-dipole interactions, it was discovered that the magnetic susceptibility of the material possessed an unexpected, slow time-dependent drift in the very low magnetic field environment in which the measurement was conducted. In the foreseeable future the sensitivity of all of these techniques will continue to improve. It is therefore important to engage in an experimental investigation of methods to suppress the magnetic susceptibility of the test mass materials.\\

A small magnetic susceptibility and a large number density of electrons and nuclei are not the only desirable features for the test masses in such experiments. It would be nice if the magnetic susceptibility of the test mass could be changed nonmagnetically with negligible disturbance of its dimensions and other relevant properties in the experiment to help isolate such systematic errors. One must also consider magnetic noise as well. If the mass is an electrical conductor, thermally-induced current fluctuations can produce magnetic fields which are now large enough to disturb sensitive magnetometers. For example, the inner layer of the magnetic shielding of Spin Exchange Relaxation Free (SERF) magnetometers based on spin-exchange optical pumping must now be chosen using an electrically-insulating ferromagnetic materials like ferrites for this reason~\cite{Kornack:2007}. For NMR-based magnetometers employing time-dependent magnetic fields, the presence of an electrical conductor produces fields from eddy currents which will disturb the measurement. For all of these reasons one would ideally want the low magnetic susceptibility material to also be an electrical insulator. Although one can introduce a superconducting shield between the test mass and the experiment, this introduces obvious practical limitations on the separation between the test mass and the polarized ensemble due to the required cryogenics. Many of the scientifically interesting spin-dependent forces of interest possess interaction ranges below the millimeter scale, which tends to make a solution like superconducting shields difficult to implement.\\

Engineering a material with no static magnetic susceptibility averaged over a macroscopic volume is a nontrivial task for various reasons. The magnetic susceptibility of many materials in practice is often determined by (magnetic) impurities, and so one essential criterion for a practical low susceptibility material is the ability to remove its magnetic impurities. Existing SERF magnetometers are already sensitive to ppb magnetic impurities held at distances as far as a few cm from the polarized gas ensemble. Perfect crystals and liquid metals come to mind as possible candidates for materials with relatively low magnetic impurities: clearly, chemical synthesis with pure materials can address this issue directly. Certain materials (quartz, sapphire, silicon, gallium etc.) can in principle at the time of this writing be obtained commercially with the required purity. In the case of sapphire, delicate measurements conducted in low-temperature sapphire resonators developed as time standards have been used to verify the absence of magnetic impurities by direct measurement~\cite{Warrick:2013}. Large crystals grown as scintillator detectors in nuclear and high energy physics might well possess small enough magnetic impurities: we are not aware of any sensitive magnetic measurements that have been conducted in these materials.\\

Even in a pure material, one always has the diamagnetic or paramagnetic contribution to the susceptibility from the electronic orbital or spin response for its specific atomic or chemical structure, and typically it is not easy to tune these values without either changing the conditions for motion of the electrons or changing the dynamics of the spin or orbital moments. Although the diamagnetic susceptibility is normally independent of temperature and the paramagnetic susceptibility is weakly temperature dependent near room temperature according to the Curie-Weiss law, one is unlikely to be able to find a preexisting material with both paramagnetic and diamagnetic contributions to the susceptibility with nearly equal magnitudes and which also nearly cancel around room temperature. The anisotropic magnetic susceptibility induced by nontrivial crystal structure further complicates this task. One solution is to mix different materials of opposite magnetic susceptibility in the right proportions to make a zero susceptibility mixture near room temperature and choose the proportions under the assumption that the susceptibility is not modified in the mixture in some nontrivial way. This condition is known as the Wiedemann's additivity law~\cite{Field:1989} in the literature. 
\begin{equation}
\chi_{\alpha \beta} m_{\alpha \beta}= \chi_{\alpha} m_{\alpha} + \chi_{\beta} m_{\beta}
\end{equation}
where $\alpha$, $\beta$ and $\alpha \beta$ denote different components and the solution/mixture.
This law should be well-satisfied for inert materials: typically susceptibility modifications upon mixture only happen from some chemical reaction or from a change in the electrical conductance. Such susceptibility changes can be used to learn about features of the relevant physics and chemistry~\cite{Trz:1982}.  Often the magnetic susceptibility changes noticeably only with a change in the thermodynamic phase of the material at a magnetic phase transition. Many such phase transitions are either first order (and therefore accompanied with a latent heat and the potential for forming metastable states) or second order (and therefore typically associated with some form of magnetic order which one does not want in the first place). The very extensive $R\&D$ done on materials with reversibly changeable magnetic properties for microelectronics typically searches for relatively large susceptibility changes in thin film magnetic materials and is unfortunately of little use for our purposes as we are typically interested so far in searching for spin-dependent forces over longer distance scales than the thickness of these films.\\  

In this paper, we report the magnetic properties of three qualitatively different classes of materials which fulfill many of these experimental requirements. We have successfully fabricated and characterized four different materials that fall under three categories: (I) solutions of paramagnetic salts and diamagnetic liquids, (II) liquid gallium-indium alloys and (II) pressed powder mixtures of tungsten and bismuth. We present a thorough discussion of the fabrication processes of these materials and show their measured magnetic susceptibilities as a function of the proportions of the different mixtures in the following sections. In all cases, the susceptibilities are a linear function of the mixture proportions as predicted by the Wiedemann law. Although this is perhaps not surprising, it is interesting to demonstrate this behavior explicitly in our case as it adds confidence in the predictability of the mixing procedure for the fabrication of zero magnetic susceptibility materials. Magnetic susceptibility $\chi$ is defined as the magnetization M (magnetic dipole moment per unit volume) induced by an applied magnetic field H.
\begin{equation}\label{eq1}
M =\chi H
\end{equation}
$\chi$ is unitless in SI system and all the measured values in this paper will be reported in the cgs units of volume magnetic susceptibility.\\ 

\section{Measurement System}

Many methods have been developed for the measurement of DC magnetic susceptibility~\cite{Philo:1977,Davis:1993,Barmet:2007} which are more sensitive and practical than the classic, widely-known Guoy technique. Our magnetic susceptibility measurements were performed using a commercial magnetic susceptibility balance (MSB) Auto (Johnson Matthey) whose working mechanism is based on the Evan's balance method~\cite{Matthey:2006}. It is advertised to operate over a sensitivity range of $0.001\times10^{-7}$ to $1.99\times10^{-4}$ volume cgs units. For reference, distilled water, a commonly-used magnetic susceptibility standard has a magnetic susceptibility of $-0.712\times10^{-6}$ volume cgs units at room temperature~\cite{Skokanova:1978}.\\

\begin{figure}
\begin{center}
\includegraphics[width=7 cm]{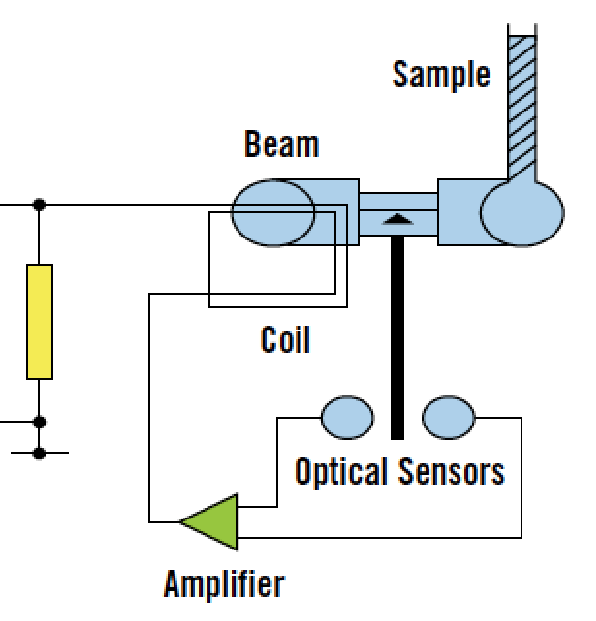}
\caption{MSB Auto schematic.}
\label{fig:fig1.}
\end{center}
\end{figure}

Fig. 1 shows MSB Auto's conceptual schematic~\cite{Matthey:2006}. It is a torsion balance containing two pair of moving permanent magnets. The magnets are positioned at the opposite ends of a beam of a torsion balance. The first pair of magnets produce a 0.5 Tesla homogenous field. When a substance with finite magnetic susceptibility is introduced in this region, it introduces a magnetic field gradient which exerts a torque on the torsion balance and torques the beam. The rotation of the balance is detected with optical sensors. A coil placed between the other pair of magnets carries a current that in turn produces a magnetic field gradient force opposing this torque. This current is proportional to the volume magnetic susceptibility of the sample. The force F acting on the sample is given by~\cite{Matthey:2006}:\\
\begin{equation}
F = \frac{\chi AH^2}{2}
\end{equation}
where $\chi$ is the volume magnetic susceptibility of the sample, A is the sample cross sectional area, H is the uniform magnetic field produced by the permanent magnets.\\

As the susceptibility measurement is conducted mechanically by a delicate torsion balance with magnetically-generated torques, it is sensitive to external perturbations like tilt, magnetic noise, and temperature variations. To minimize external magnetic field variations, we housed the device inside two concentric mumetal magnetic shields. The space between these two shields is filled with polyethylene beads to increase the heat capacity of the system and attenuate external thermal perturbations, and the whole apparatus is housed in external thermal insulation. A fluxgate magnetometer with mG sensitivity monitors the residual stray field inside the mumetal shield. The balance rests on a homemade kinematic mount on electrically insulating ceramic balls and is supported on a marble table of the type commercially produced for sensitive weight balances. A double axis tilt sensor (Aositilt EZ 5000) with 0.1 mrad sensitivity and a resistance thermometer with mK resolution recorded any change in the tilt angles and temperature during the measurement. We found that the balance would operate reproducibly at its sensitivity limit with temperature variations below 10 mK, magnetic field variations under 10 mG and tilt variations under 0.1 mrad. The magnetic susceptibility value was measured every second and all the measurements were averaged over a period of one minute. 

\section{I. Paramagnetic salt and diamagnetic liquid solutions}

A weighted concentration of paramagnetic salt and diamagnetic liquid can be mixed to form a solution with magnetic susceptibility theoretically close to zero given by Wiedemann's additivity law~\cite{Matthey:2006} as shown in Eq. (3). In our case, we chose to develop two different salt solutions: (a) manganese chloride ($MnCl_{2}$) and water (b) chromium(III) acetylacetonate ($C_{15}H_{21}Cr_{6}$) and dichloromethane (DCM)($CH_{2}Cl_{2}$). $MnCl_{2}$'s paramagnetic susceptibility is large enough to make a nominally zero-susceptibility mixture with water in small enough concentrations to avoid precipitation at room temperature~\cite{Andres:1976, Barmet:2007}. $MnCl_{2}$ was acquired from Fisher Scientific and mixed with pure water to form different concentrations of $MnCl_{2}$ solution by weight. These solutions were filled in a NMR tube and characterized using the MSB Auto. The magnetic susceptibility of the empty NMR tube was recorded first before filling it with the solutions. Corrections for the tilt and magnetic field variations were very small thus unnecessary. One must correct for the paramagnetic susceptibility of the oxygen in the air at this sensitivity, and since the paramagnetic susceptibility is temperature dependent, we extrapolated all of the measurement results to a common temperature of 296.93K. The corrections were calculated using the known temperature dependence of the susceptibilities of the pure elements near room temperature~\cite{Suzuki:1971, Uetake:2000} when applicable and weighting them appropriately.\\ 

Table 1 contains the different concentration of $MnCl_{2}$ solutions vs. their measured magnetic susceptibility plotted in Fig. 2. From this linear plot, it was confirmed that a 1\%\ $MnCl_{2}$ solution had nearly zero magnetic susceptibility. And this solution was used as one of the nonmagnetic test masses in a recent spin-dependent interaction experiment~\cite{Chu:2013} that set new and improved limits on monopole-dipole interactions in the range from $10^{-3}$ to $10^{-4}$m. The error bar for solutions in Table 1 is the approximate average error of the MSB Auto. Rest of the solutions and mixtures discussed later contain their statistical error bars.\\

\begin{table}[t]
\centering
\begin{tabular}{ c || c }
\textbf{$MnCl_{2}$}\boldmath$\%$ & \boldmath$\chi\times10^{-6}$ \textbf{(cgs volume)} \\
\hline
0$\%$ & $-0.716 \pm 0.094$\\
0.5$\%$ & $-0.366 \pm 0.094$\\
1$\%$ & $-0.012 \pm 0.094$\\
1.5$\%$ & $0.344 \pm 0.094$\\
2$\%$ & $0.686 \pm 0.094$\\
\hline
\end{tabular}
\caption{Volume magnetic susceptibility values of $MnCl_{2}$ and water solution according to weight percent of $MnCl_{2}$. The errors are dominated by systematic error and the source is explained in the text.}
\end{table}

\begin{figure}
\begin{center}
\includegraphics[width=7 cm]{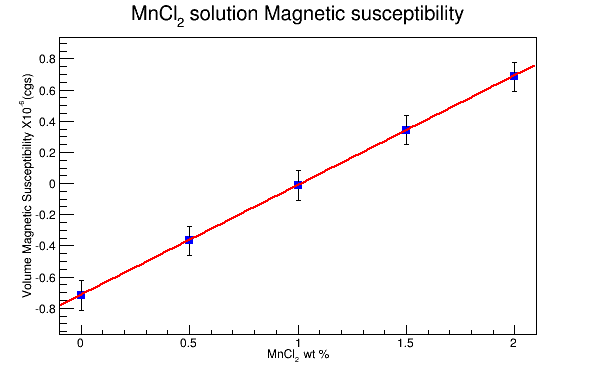}
\caption{volume magnetic susceptibility of $MnCl_{2}$ and water solution vs. their $MnCl_{2}$ wt. percent composition.}
\label{fig:fig2.}
\end{center}
\end{figure}

Similarly, paramagnetic chromium acetylacetonate was mixed with common diamagnetic organic solvent DCM to form very low magnetic susceptibility solutions~\cite{Jarrett:1957, Sharma:1986}. 97$\%$ chromium(III) acetylacetonate and DCM were obtained from ACROS Organics and EMD Performance materials respectively. Magnetic susceptibility measurements for these organic solutions as a function of the Chromium(III) acetylacetonate concentration are given in Table 2. The relationship between the measured magnetic susceptibility and different concentrations is shown in Fig. 3. The solution with the lowest magnetic susceptibility was temporarily used in the spin-dependent interaction experiment of Ref.~\cite{Chu:2013} as the source of nonmagnetic test mass. Although the use of this solution was discontinued in this experiment after it dissolved the very thin teflon film membrane which separated the liquid from the glass cell of polarized nuclei, it may be an interesting choice for other experiments which require a zero magnetic susceptibility room temperature liquid. 

\begin{table}[t]
\centering
\begin{tabular}{ c || c }
\textbf{$C_{15}H_{21}Cr_{6}$}\boldmath$\%$ & \boldmath$\chi\times10^{-6}$ \textbf{(cgs volume)} \\
\hline
0$\%$ & $-0.715 \pm 0.056$\\
1.5$\%$ & $-0.304 \pm 0.036$\\
3$\%$ & $-0.01 \pm 0.061$\\
4.7$\%$ & $0.351 \pm 0.018$\\
6.2$\%$ & $0.786 \pm 0.101$\\
\hline
\end{tabular}
\caption{Volume magnetic susceptibility values of solution of dichloro methane (DCM)($CH_{2}Cl_{2}$) and chromium (III) acetylacetonate ($C_{15}H_{21}Cr_{6}$) according to weight percent of chromium acetylacetonate. The errors are dominated by statistical error and come from the noise in the magnetic susceptibility meter data.}
\end{table}

\begin{figure}
\begin{center}
\includegraphics[width=7 cm]{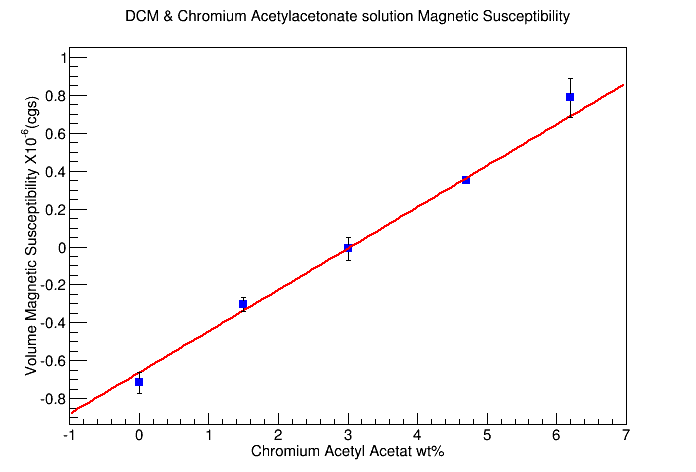}
\caption{Volume magnetic susceptibility of chromium (III) acetylacetonate ($C_{15}H_{21}Cr_{6}$) and dichloromethane (DCM)($CH_{2}Cl_{2}$) vs. their chromium acetylacetonate wt. percent composition. The errors are dominated by statistical error and come from the noise in the magnetic susceptibility meter data.}
\label{fig:fig2.}
\end{center}
\end{figure}

\section{II. Gallium-Indium Alloys}

Gallium is the least magnetic element. Its orientation-averaged magnetic susceptibility $\chi$ in the solid phase of $-0.248 10^{-6} \frac{M}{g}$ (cgs mass susceptibility units)~\cite{Pashaev:1973} drops to the amazingly small value of $0.002\times10^{-6} \frac{M}{g}$ in the liquid phase, two orders of magnitude smaller. This anomalously small value of magnetic susceptibility seems to come from an accidental cancellation of the atomic susceptibility with that from the free electrons in the liquid metal~\cite{Aleksandrov:1961}. As a liquid metal with a low vapor pressure, it has the potential to be produced with very high purity: gallium with ppb magnetic impurities is commercially available, and research projects in progress foresee the possibility to produce gallium with magnetic impurities at even lower levels. In the absence of an oxide layer (which can be prevented in an inert atmosphere and mitigated with mild acids~\cite{Xu:2012, Kim:2008}), the surface tension of liquid gallium causes it to bead up on almost all surfaces, which is a good property if one is trying to get the gallium as close to an ensemble of polarized nuclei or electrons without interacting with the surface of the boundary that separates them. Unfortunately liquid gallium has two unfavorable properties for our application. As a metal, it is a source of magnetic noise from thermal currents. In addition, its melting point is slightly higher than room temperature (29.78$^{\circ}$C)~\cite{Xu:2012} which can be inconvenient for many precision experiments.\\

The main motivation for this part of the work was therefore to see if we could address the latter issue by mixing gallium with another material which can also be purified in principle to form a mixture that is both free of magnetic impurities, liquid at room temperature, and with a low magnetic susceptibility. Indium is a low $\chi$ element with a high nucleon density with many very similar physical properties to gallium. Indium is solid at room temperature and has $\chi$ of $-0.112\times10^{-6} \frac{M}{g}$ cgs mass magnetic susceptibility~\cite{Dumke:1983}. It can be liquefied without difficulty and therefore can be produced with minimal impurities. Furthermore, gallium-indium alloys are liquid at and below room temperature over a broad range of indium concentrations~\cite{Suzuki:1971,Pashaev:1973}. We also wanted to check to see if the magnetic susceptibility of the liquid gallium-indium mixtures is the same as the weighted sum of the magnetic susceptibilities of the components.\\ 

We used 4N-purity gallium and indium from American Elements. All the tools used to contain or manipulate these alloys were cleaned very thoroughly first with soap and distilled water and in 10 $\%$ hydrochloric acid (when compatible), and washed finally with alcohol and acetone. The melting point of indium is 156$^{\circ}$C~\cite{Xu:2012}. The appropriate proportions of these two metals were weighed and put in a borosilicate glass beaker (VMR International) on top of a hot plate set to 164$^{\circ}$C. This temperature ensured complete melting of indium along with gallium and minimized indium oxide formation which rapidly increases with higher temperature~\cite{Kim:2007}. A teflon covered magnetic stir rod was used to homogenize this alloy while heating. The oxide layer readily stuck to the surface of the glass as the alloy was stirred leaving the alloy silvery and clean. After about 15 minutes, it was left to cool down for a few minutes before drawing it into a borosilicate glass syringe with all nonmetallic parts and a needle made of Peek plastic tubing to ensure that no metallic/magnetic impurities were accidentally introduced into the alloy. The mixtures were all syringed into glass NMR tubes for characterization.\\

\begin{table}[t]
\centering
\begin{tabular}{ c || c }
\textbf{In wt.}\boldmath$\%$ & \boldmath$\chi + \Delta \chi\times10^{-6}$ \textbf{(cgs volume)} \\
\hline
0$\%$ & $-0.002 \pm 0.013$\\
5$\%$ & $-0.06 \pm 0.045$\\
10$\%$ & $-0.095 \pm 0.025$\\
12$\%$ & $-0.106 \pm 0.018$\\
13.4$\%$ & $-0.117 \pm 0.014$\\
16.5$\%$ & $-0.126 \pm 0.017$\\
\hline
\end{tabular}
\caption{Volume magnetic susceptibility of gallium-indium alloys along with the errors according to the indium atomic \% composition. The errors come from the noise of the susceptibility meter.}
\end{table}

Table 3 shows the measured average volume magnetic susceptibilities of different gallium-indium alloy mixtures in cgs units taken over three different measurements. Their magnetic susceptibilities are as expected based on the weighted sum of the susceptibilities of the elements. \\
\begin{figure}
\begin{center}
\includegraphics[width=7 cm]{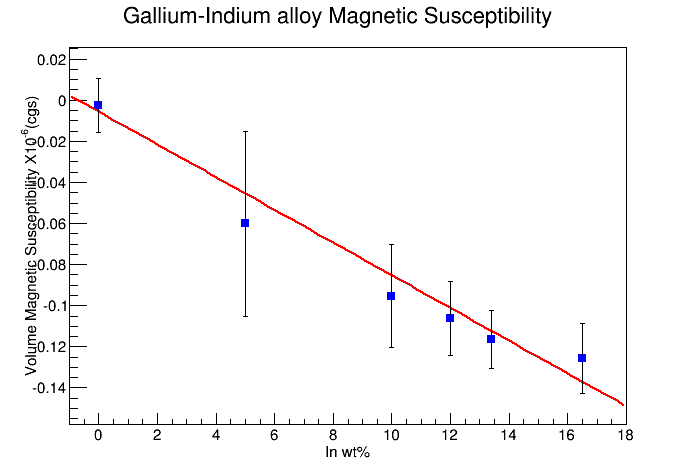}
\caption{volume magnetic susceptibility of gallium-indium alloys vs. their indium atomic percent composition.}
\label{fig:fig3.}
\end{center}
\end{figure}

Fig. 4 shows the plot of volume magnetic susceptibility vs. temperature for these alloys. The magnetic susceptibility is a linear function of the indium percent in gallium-indium alloys over the range of indium fractions measured within experimental errors. The results are consistent with earlier measurements~\cite{Suzuki:1971} conducted on both pure gallium and the gallium-indium alloy at the eutectic mixture point of 16.5\% indium. 

\section{III. Tungsten Bismuth metallic mixtures}
Materials in powder form with opposite magnetic susceptibilities can also be mixed together to yield very low volume-averaged magnetic susceptibilities~\cite{Bulatowicz:2014}. Such powders could be sintered in principle (if needed) and made into a solid, which might be preferable to a liquid as a nonmagnetic mass source in some experiments. We used 99.5\%\ 200-325 mesh tungsten (paramagnetic) and bismuth (diamagnetic) powders from Alfa Aesar. Different percent composition mixtures (by weight) of these two were made and characterized using the MSB Auto balance. The preliminary values of measured susceptibilities and the percent composition of these mixtures are reported in Table 4  and Fig. 5 shows the magnetic susceptibility of these mixtures plotted against the wt \%\ bismuth content.\\

\begin{table}[t]
\centering
\begin{tabular}{ c || c }
\textbf{Bi wt.}\boldmath$\%$ & \boldmath$\chi \times10^{-6}$ \textbf{(cgs volume)} \\
\hline
19$\%$ & $6.24 \pm 0.091$\\
34$\%$ & $2.40 \pm 0.063$\\
43$\%$ & $1.42 \pm 0.017$\\
50$\%$ & $0.006 \pm 0.063$\\
51.2$\%$ & $-0.06 \pm 0.058$\\
\hline
\end{tabular}
\caption{preliminary results of volume magnetic susceptibility $\chi$ measurement of tungsten-bismuth powder mixtures according to the constituent wt. $\%$ of bismuth}
\end{table}

Initial effort to pressing the powder into pellets involved the use of Wabash hydraulic hot-press, pressing the mixture to 40,000lb. Bismuth has much lower melting point of 150$^{\circ}$C to 200$^{\circ}$C as oppossed to 3400$^{\circ}$C for tungsten, thus, we expected liquid-state sintering effect. The solids that formed with this procedure were delicate but compact enough to stay together. The pressing procedure could be further enhanced by using higher pressure and temperature to produce more durable solids that do not break easily. 

\begin{figure}
\begin{center}
\includegraphics[width=7 cm]{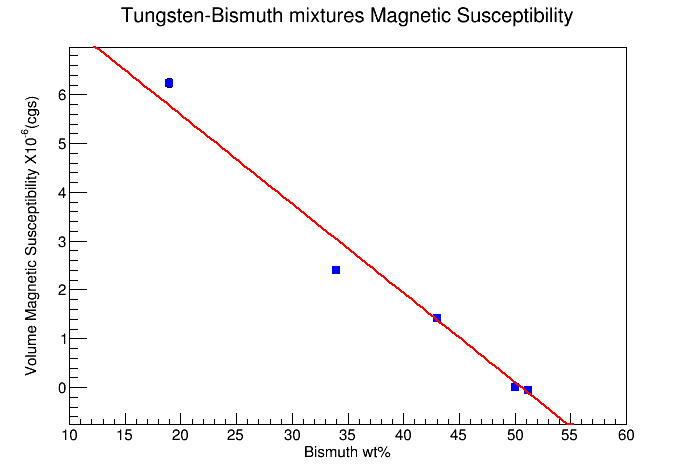}
\caption{volume magnetic susceptibility of tungsten-bismuth metallic mixtures vs. their bismuth \%\ wt. composition. Note some of the error bars are too small to see.}
\label{fig:fig5.}
\end{center}
\end{figure}

\section{Conclusion}

We have fabricated and characterized a number of solids and liquids with very low magnetic susceptibility at room temperature. These materials have the potential to be used as the source of unpolarized/nonmagnetic mass in experiments where magnetic susceptibility of the masses is a potential source of systematic error. Some of them also have high nucleon density which is desirable for some experiments. We anticipate that further measures to protect against magnetic susceptibility systematic errors in precision experiments will need to be employed in the future. Fortunately there is much room for improvement. Some strategies which suggest themselves and have been employed in previous work or are contemplated for future experiments include, but are not limited to, (1) the use of co-magnetometers for the polarized species, with one species effectively measuring the magnetic field shift from the test mass and the other species searching for the exotic spin-dependent interaction~\cite{Tullney:2013, Bulatowicz:2013}, (2) construction of segmented test masses composed of two different materials so that the oscillation of the particle density occurs at a frequency $n\omega$ for a spinning frequency $\omega$ of the object~\cite{Geraci:2010}, (3) superconducting magnetic shields interposed between the mass and the polarized species~\cite{Geraci:2010}, (4) exploitation of the anisotropy of the magnetic susceptibility of crystal test masses to vary the susceptibility at constant particle density, (5) use of the temperature dependence of the paramagnetic susceptibility near a phase transition to adjust the test mass susceptibility, (6) use of the known functional form of the fifth force interaction along with a readout scheme which can exploit this dependence. In some cases, it may be necessary to further reduce the magnetic impurities of solid materials by special high purity sample preparation methods like those which are already used to produce silicon, quartz, and sapphire with ppb magnetic impurities. One can imagine amplifying the contribution of magnetic impurities to the susceptibility by immersing the test mass in a large external magnetic field in an attempt to magnetize any existing impurities above detection threshold. Obviously, most technical advances in magnetometry can be adapted in principle to perform improved magnetic susceptibility measurements.\\

\begin{acknowledgements}

This work was supported in part by NSF PHY-1068712 and NSF PHY-1306942. All of the authors acknowledge support from the Indiana University Center for Spacetime Symmetries. L Dennis acknowledges support through the NSF Research
Experiences for Undergraduates program NSF PHY-1156540.\\

\end{acknowledgements}

\end{document}